%% file: CahoyMain.tex
\documentclass[smallextended]{svjour3}     
\usepackage{graphicx}
\usepackage{fix-cm}

\usepackage{amssymb}
\usepackage{amsmath}
\usepackage{comment}
\usepackage{bm}
\usepackage{multirow}

\numberwithin{equation}{section}
\allowdisplaybreaks

\begin{document}

\title{An estimation procedure for the Linnik distribution}


\author{Dexter O. Cahoy}


\institute{Dexter O. Cahoy  $( \boxtimes )$ \at
       Program of Mathematics and Statistics\\
       College of Engineering and Science\\
		Louisiana Tech University, USA\\
        Tel: +1 318 257 3529\\
        Fax: +1 318 257 2182\\
         \email{dcahoy@latech.edu}
}

\date{Received: 09-22-2010  / Accepted: date}

\maketitle

\begin{abstract}
        We propose  estimators for the parameters of the   Linnik L$(\alpha,\gamma)$ distribution.  The estimators are derived from the moments of the log-transformed Linnik distributed random variable, and are shown to be asymptotically unbiased. The estimation algorithm  is computationally simple and less restrictive. 
Our procedure is also tested  using simulated data.

\end{abstract}

\input{s1.tex}
\input{s2.tex}

\input{s3.tex}

\begin{acknowledgements}
The author is grateful to the Editor and the Reviewer for their insightful suggestions and comments   that
significantly improved the article. 
\end{acknowledgements}


%
%

\end{document}

%% file: s1.tex
\section{Introduction} \label{a}

In recent years, the Linnik  L$(\alpha, \gamma)$  distribution of \cite{lin63} has gained popularity from researchers in many scientific areas. For instance, it has been used to model  random phenomena  in  finance (e.g., S\&P index) \cite{koz99,koz01}. In addition, \cite{dev90,koz01,kao96,lin98a,pake98} (and the references therein)  studied the   L$(\alpha, \gamma)$  probability density function with characteristic function (ch.f.)
\begin{equation*}
\psi(\lambda) = \left( 1 +  |\gamma \lambda|^\alpha  \right)^{-1}, \tag{1.1}    \label{1.1}
\end{equation*}
where  $\gamma>0$ is the scale parameter, $\lambda \in \mathbb{R}$, and $0 < \alpha \leq 2$.  In particular, the probability density and the cumulative distribution functions for a L$(\alpha, 1)$ distributed random variable  are
\[
f(x)= \frac{\sin (\pi \alpha/2)}{\pi} \int_0^\infty \frac{y^\alpha \exp (-xy)}{y^{2\alpha} +2y^\alpha \cos (\alpha \pi/2) +1} dy,
\]
and
\[
F(x)=1- \frac{\sin (\pi \alpha/2)}{\pi} \int_0^\infty \frac{y^{\alpha-1} \exp (-xy)}{y^{2\alpha} +2y^\alpha \cos (\alpha \pi/2) +1} dy, \qquad x >0,
\]
correspondingly. Moreover, \cite{koz01} constructed the following structural representation of a L$(\alpha, \gamma)$  distributed random variable $L$:
\begin{equation*}
L\stackrel{d}{=} \gamma D R^{1/\alpha},  \tag{1.2}    \label{1.2}
\end{equation*}
where $D$ has the standard Laplace distribution (with location 0 and scale 1), and $R$ has the density function
\[
f_R(r)= \frac{\sin (\pi \rho)}{\rho \pi \left[ r^2 +2r\cos(\rho \pi) +1 \right]}, \quad 0< \rho <1, \; r >0.
\] 
It is known that the  L$(\alpha, \gamma)$  distribution is geometric stable.

The parameter estimation problem for $\alpha$ when $\gamma=1$ was addressed by \cite{a92} using the methods of \cite{lap75,phl75,press72}. Then \cite{jrt99} adopted  \cite{press72}'s technique to estimate the parameters  $\alpha$ and $\gamma$ of the Linnik L$(\alpha, \gamma)$ distribution.  Furthermore,  \cite{jrt99} constructed estimators that require choosing values of $\lambda$, $\lambda_i$'s say such that $\lambda_i \in \mathbb{R} \setminus {0}, \; i=1,\ldots,b$, and $\sum_{j=1}^b \left( \log |\lambda_j| - c \right)^2>0$, where $c= (1/b)\sum_{j=1}^b \log |\lambda_j|$.
 More importantly, \cite{jrt99} deduced that $\lambda_i$'s should be restricted to a region where $\log \left(  |\hat{\psi} (\lambda)| - 1 \right) $ is `linear' with respect to $\log |\lambda|$ to obtain satisfactory results. Apparently, satisfying the above restrictions is not straightforward in practice. Note that $\hat{\psi} (\lambda)$  is the method-of-moments estimator of the characteristic function (\ref{1.1}). Moreover, they showed the asymptotic normality of their point estimators, i.e.,
\[
\sqrt{n}\left(
  \begin{array}{c}
    \hat{\alpha}_P-  \alpha \\
    \hat{\gamma}_P - \gamma \\
  \end{array}
\right) \stackrel{d}{\longrightarrow}  \textsl{N} \left[\bm{0} , \textbf{W}_P \right]
\]
as $n \to \infty$, where the entries of the covariance matrix $\textbf{W}_P=\left( \textbf{W}_{P_{ij}} \right)$ are defined as
\[
\textbf{W}_{P_{11}} =
         \textbf{u}^\text{T}\textbf{Wu}, \qquad \qquad  \textbf{W}_{P_{12}}=\textbf{W}_{P_{21}} =  \frac{\gamma}{\alpha}\left( \textbf{1}/b - \textbf{u}(c + \log \gamma ) \right)^\text{T}\textbf{W}\textbf{u},
\]
\[
\textbf{W}_{P_{22}} =\left( \frac{\gamma}{\alpha} \right)^2\left( \textbf{1}/b - \textbf{u}(c + \log \gamma ) \right)^\text{T}\textbf{W} \left( \textbf{1}/b - \textbf{u} ( c + \log \gamma ) \right),
\]
$ \textbf{W} = \left(w_i w_j w_{ij} \right)$, $w_i = \left(\psi (\lambda_i)^{-2} \right)/ |\gamma \lambda_i|^\alpha $, $\textbf{1}=(1,\ldots,1)^\text{T}$,  $ u_i=(\log |\lambda_i |-c)/ \sum_{j]1}^b \left(\log |\lambda_j |-c \right)^2,$ \textbf{u}$^\text{T} =(u_1,\ldots,u_b)$, and  $w_{ij} = [ \psi (\lambda_i + \lambda_j) +   \psi (\lambda_i - \lambda_j)$ \linebreak $-2  \psi (\lambda_i)  \psi (\lambda_j)]/2 $. The point estimators of $\alpha$ and $\lambda$ are given by
\[
\hat{\alpha}_P= \sum\limits_{j=1}^b \log \left( |\hat{\psi}|^{-1} - 1 \right)u_j
\]
and
\[
\hat{\gamma}_P=\exp \left\lbrace \frac{1}{b \hat{\alpha}_P}  \sum\limits_{j=1}^b \log \left( |\hat{\psi}|^{-1} - 1 \right) - c \right\rbrace,
\]
correspondingly. Observe that one can always use the tail-index estimators of \cite{hnr80,hall82,hil75} for the parameter $\alpha$ as well. However, these methods impose restrictions or use a portion of the data only  making them less  efficient.

Similarly, \cite{koz01} suggested the fractional moment estimators, i.e., choose constants $q_1, q_2 <\alpha$, and calculate the parameter estimates $\hat{\alpha}_K$ and $\hat{\gamma}_K$ by solving the following system of two non-linear equations:
\[
\hat{\mu}_{|L|^{q_j}} = \widehat{\mathbf{E}|L|^{q_j}} =  \frac{\pi q_j(1-q_j) \hat{\gamma}_K^{q_j} }{ \hat{\alpha}_K \Gamma (2-q_j) \sin \left(\pi q_j/\hat{\alpha}_K \right) \cos \left( \pi q_j/2\right) }, \qquad  j=1,2.
\]
Clearly, the pre-selection of appropriate values  requires  information about the true or unknown parameter $\alpha$ a priori, which is not feasible in practice.  As a direct consequence, it is expected that the above estimators will perform poorly when the restrictions are violated. Notice also that the point estimators of \cite{jrt99}  have complicated expressions while \cite{koz01}'s is computationally involved.  It is mainly these drawbacks  that stimulate us to construct a simple estimation algorithm  that uses all the available information possible,  and avoids the above limitations.

The main goal of this paper is to propose estimators  for the  parameters $\alpha$ and $\gamma$ of the Linnik L$(\alpha, \gamma)$ distribution. The rest of the paper is organized  as follows: In Section 2, we  review a structural representation of a  Linnik L$(\alpha, \gamma)$  random variable via symmetric stable. In Section  3,  we derive the method-of-moments estimators. In Section 4, the asymptotic normality of the estimators are then shown. Empirical test results are given in Section 5. Section 6 discusses key points and extensions of our study.

%% file: s2.tex
\section{Stable representation of the random variable  $L$}

For the sake of completeness, we review the stable representation of the random variable $L$ in (\ref{1.1}) as derived by \cite{dev90}.

\noindent \newtheorem{thm1}{Result}
\begin{thm1}
Let $0< \alpha \leq 2$, and  the scale parameter $\gamma > 0$. Then
\begin{equation*}
L \stackrel{d}{=} \gamma Z^{ 1 /\alpha } S  \tag{2.1} \label{2.1}
\end{equation*}
where $S$  is a symmetric stable distributed random variable with the characteristic function $\psi_S(\lambda)=\exp (-| \lambda|^\alpha)$, and $Z$ is an exponentially (with scale 1) and independently distributed random variable.
\end{thm1}

An expression for the $q$th fractional moment can be straightforwardly derived from the above result, and is given below.

\noindent \newtheorem{cor1}{Proposition}
\begin{cor1}
Let $0< \alpha \leq 2$, and  the scale parameter $\gamma > 0$. Then
\[ 
\mathbf{E} |L|^q = \frac{\pi q \gamma^q }{\alpha \sin (\pi q/ \alpha) \cos (\pi q/2) \Gamma (1-q)}, \qquad 0<q < \alpha.  
\]
\end{cor1}
\noindent \textbf{Proof}. The  $q$th fractional moment is

\begin{align} \notag
\mathbf{E} |L|^q  &= \gamma^q \mathbf{E} |S|^q \mathbf{E}(Z^{q / \alpha}).  \notag\\ \notag
\end{align}

Using the $q$th fractional moment of the symmetric stable random variable $S$ (see \cite{bkkuz04} )
\begin{equation*}
w_\alpha(q)= \mathbf{E} |S|^q  = \frac{\Gamma (1-q/\alpha)}{\cos (q \pi/2) \Gamma (1-q)},  \tag{2.2} \label{2.2}
\end{equation*}
we have
\begin{align} \notag
\mathbf{E} |L|^q          &= \gamma^q \left( \frac{\Gamma (1-q/\alpha)}{\cos (q \pi/2) \Gamma (1-q)} \right) \Gamma (1 + q / \alpha). \notag\\ \notag
\end{align}
Substituting
\[
\Gamma ( 1 - q/\alpha ) = \frac{\pi}{\Gamma( q /\alpha) \sin (\pi q/\alpha )}, \qquad\text{and} \qquad   \Gamma ( 1 + q/\alpha ) = \frac{q}{\alpha} \Gamma( q /\alpha)
\]
into the preceding equation completes the proof. Observe that the $q$th fractional moment above has a different and simpler form  in comparison with \cite{koz01}.

\section{Method-of-Moments (MoM) estimation}

Applying the log transformation to the absolute value of the random variable $L$ given in (\ref{2.1}), we get
\begin{equation*}
L^{'} \stackrel{d}{=} \log (\gamma ) + \frac{1}{\alpha}Z^{'} + S^{'},  \tag{3.1} \label{3.1}
\end{equation*}
where $L^{'}= \log (|L|)$, $Z^{'}=\log (Z)$, and $S^{'}=\log (|S|)$. From \cite{cuw10}, we have
\[
\mathbf{E}(Z^{'})=-\mathbb{C}, \qquad \mathbf{E}
\left( Z^{'} \right)^2=\mathbb{C}^2+\frac{\pi^2}{6},
\]
\[
\mathbf{E}\left(  Z^{'} \right)^3 = -
\mathbb{C}^3-\frac{\mathbb{C}\pi^2}{2}-2\zeta (3), \qquad
\mathbf{E}\left( Z^{'}\right)^4 =
\mathbb{C}^2\left(\mathbb{C}^2+\pi^2\right) +
\frac{3\pi^4}{20}+8\mathbb{C}\zeta (3),
\]
where $\mathbb{C}\simeq 0.5772156649015328606065$ is the Euler's constant. Recall  the following formula from \cite{bkkuz04} and \cite{zol86} for the higher log-moments of $|S|$:
\[
\mathbf{E}\left( S^{'} \right)^k=\left(d^kw_\alpha (q)/dq^k
\right)\big|_{q=0},
\]
where $w_\alpha (q)$ is defined in (\ref{2.2}).  We now need to find the power series expansion of $ w_\alpha (q)$. But this turns out to be easier if we first
expand
\[
\log w_\alpha (q) = \log \Gamma (1-q/ \alpha) - \log \Gamma (1-q) -  \log \cos (q \pi / 2)
\]
into a power series. Using the well-known expansions
\[
\log \Gamma (1-\theta) = \mathbb{C} \theta + \sum
\limits_{k=2}^\infty \frac{ \zeta (k)}{k} \theta^k ,
\]
and
\[
\log \cos (\theta) = - \theta^2/2 + O(\theta^4),
\]
we get
\[
\log w_\alpha (q) = \mathbb{C}\left( \frac{1}{\alpha}-1\right)q +
\frac{\pi^2 \left( \alpha^2 + 2 \right)}{24 \alpha^2} q^2 +
\frac{\left( 1- \alpha^3 \right) \zeta (3)}{3 \alpha^3}q^3  + \frac{\left(8 +7\alpha^4 \right)\pi^4}{2880 \alpha^4}q^4  + O(q^5),
\]
where  $\zeta (\theta)$ is the Riemann zeta function evaluated at $\theta$. This implies that
\begin{align}
w_\alpha (q) &=1 + \mathbb{C}\left( \frac{1}{\alpha}-1\right)q + \bigg[
\frac{12 \mathbb{C}^2(\alpha-1)^2 + \left( \alpha^2 + 2 \right)  \pi^2 }{24 \alpha^2}\bigg] q^2 \notag \\
& + \bigg[\frac{(1-\alpha) \left(4(\alpha-1)^2\mathbb{C}^3 + (\alpha^2+2)\mathbb{C}\pi^2 + 8 (\alpha^2 + \alpha +1) \zeta (3)\right)}{24 \alpha^3}\bigg]q^3 \notag \\
& + \frac{1}{5760 \alpha^4}\bigg[ 240 (\alpha-1)^4\mathbb{C}^4 + 120 (\alpha-1)^2( \alpha^2+2)\mathbb{C}^2 \pi^2  \notag \\
& + (19\alpha^4 +20 \alpha^2 +36)\pi^4 + 1920 ( \alpha-1)^2(\alpha^2 + \alpha +1) \mathbb{C} \zeta (3) \bigg]q^4 +O(q^5). \notag
\end{align}
The $k$th log-moment is simply the coefficient of the term
$q^k/k!$ in the above power series expansion. In particular, the first four integer-order log-moments can be easily  deduced as
\[
\mathbf{E}\left(  S_\alpha^{'} \right) = \mathbb{C}\left( \frac{1}{\alpha}-1\right), \qquad  \mathbf{E} \left( S_\alpha^{'}\right)^2=
\frac{12 \mathbb{C}^2(\alpha-1)^2 + \left( \alpha^2 + 2 \right)  \pi^2 }{12 \alpha^2},
\]
\[
\mathbf{E} \left(S_\alpha^{'}\right)^3= \frac{(1-\alpha) \left(4(\alpha-1)^2\mathbb{C}^3 + (\alpha^2+2)\mathbb{C}\pi^2 + 8 (\alpha^2 + \alpha +1) \zeta (3)\right)}{4 \alpha^3} , \quad \text{and}
\]
\[
\mathbf{E} \left(S_\alpha^{'}\right)^4 = \frac{1}{240 \alpha^4}\bigg[ 240 (\alpha-1)^4\mathbb{C}^4 + 120 (\alpha-1)^2( \alpha^2+2)\mathbb{C}^2 \pi^2  \notag
\]
\[
\qquad + \; (19\alpha^4 +20 \alpha^2 +36)\pi^4 + 1920 ( \alpha-1)^2(\alpha^2 + \alpha +1) \mathbb{C} \zeta (3) \bigg].
\]
Using the above moments  and  the structural equality (\ref{3.1}), we get the mean and variance
\begin{equation*}
\mu_{L^{'}}= \log (\gamma ) -\mathbb{C}, \quad \text{and} \quad \sigma_{L^{'}}^2= \frac{\pi^2 (\alpha^2+4)}{12 \alpha^2}, \tag{3.2} \label{3.2}
\end{equation*}
respectively. Hence, we have the MoM estimators of $\alpha$ and $\gamma$ as
\begin{equation*}
\hat{\alpha}=\frac{\pi}{\sqrt{3\left( \hat{\sigma}_{L^{'}}^2 -\pi^2/12 \right)}} , \qquad \text{and} \qquad \hat{\gamma}= \exp ( \hat{\mu}_{L^{'}}  +  \mathbb{C}),  \tag{3.3} \label{3.3}
\end{equation*}
correspondingly.  Moreover, a similar calculation gives the third and fourth central moments of $S^{'}$ as
\[
\mu_3^{'}=\mathbf{E} \left( L^{'} - \mu_{L^{'}}\right)^3= -2\zeta (3),
\]
and
\[
\mu_4^{'}=\mathbf{E} \left( L^{'} - \mu_{L^{'}}\right)^4= \frac{\pi^4(19\alpha^4 + 40 \alpha^2 + 112)}{240\alpha^4}, 
\] 
which are useful in the derivation of the interval estimates given in the next section.

%% file: s3.tex
\section{Asymptotic normality of the estimators $\hat \alpha$ and $\hat \gamma$}
We will now show that the estimators  (\ref{3.3}) of $\alpha$ and $\gamma$ are asymptotically normal. Let
\[
\hat{\mu}_{L^{'}}= \overline{L^{'}} = \frac{\sum \limits_{j=1}^n L_j^{'}}{n} \quad
\text{and}\quad \hat{\sigma}_{L^{'}}^2= \frac{\sum
\limits_{j=1}^n \left(L_j^{'}-\overline{L^{'}}\right)^2}{n}.
\]
Then the following weak convergence holds (see \cite{fer96}), i.e.,
\[
\sqrt{n}\left(
  \begin{array}{c}
    \hat{\mu}_{L^{'}}-\mu_{L^{'}} \\
    \hat{\sigma}_{L^{'}}^2 - \sigma_{L^{'}}^2  \\
  \end{array}
\right) \stackrel{d}{\longrightarrow}  \textsl{N} \left[\bm{0} , \bf{\Sigma} \right]
\]
as $n \to \infty$, where the covariance matrix $\bf{\Sigma}$ is defined as
\[
\bf{\Sigma} = \left(
       \begin{array}{cc}
         \sigma_{L^{'}}^2 & \mu_3^{'} \\
         \mu_3^{'} & \mu_4^{'}-\sigma_{L^{'}}^4 \\
       \end{array}
     \right),
\]
$\mu_3^{'}, \mu_4^{'}$, and $\sigma_{L^{'}}^2$ are given in Section 3. Using a standard result on asymptotic theory, the two-dimensional Central Limit Theorem above implies that
\[
\sqrt{n}\big(\textbf{g}(\hat{\bm{\theta}}_n)-\textbf{g}(\bm{\theta})\big)\stackrel{d}{\to} \textsl{N}\left(0,\; \bm{\dot{\textbf{g}}}(\bm{\theta})^{\text{T}}\bf{\Sigma}\bf{\dot{g}}(\bm{\theta})\right),
\]
where  $\hat{\bm{\theta}}_n=(\hat{\mu}_{L^{'}},  \hat{\sigma}_{L^{'}}^2)^\text{T},  \bf{g}$ is a mapping from $\mathbb{R}^2 \to\mathbb{R}$,  and   $\bm{\dot{\textbf{g}}}(\bf{x})$ is continuous in a neighborhood of $\bm{\theta} \in \mathbb{R}^2$. We now apply this result to the consistent estimator of $\gamma$.
Letting
\[
\textbf{g}(\mu_{L^{'}},\sigma_{L^{'}}^2) = \exp \left( \mu_{L^{'}}+\mathbb{C} \right).
\]
Then the gradient becomes
\[
\bm{\dot{\textbf{g}}}(\mu_{L^{'}},\sigma_{L^{'}}^2)= \left(
                                         \begin{array}{c}
                                            \exp \left( \mu_{L^{'}}+\mathbb{C} \right) \\
                                            0   \\
                                         \end{array}
                                        \right).
\]
This implies that
\[
\sqrt{n}\big( \hat{\gamma}-\gamma\big) \stackrel{d}{\longrightarrow}
\textsl{N} \left[0,\; \sigma_{\gamma}^2 \right],
\]
where
\begin{align}
\sigma_{\gamma}^2 &= \bm{\dot{\textbf{g}}}(\mu_{L^{'}},\sigma_{L^{'}}^2)^{\text{T}} \left(
       \begin{array}{cc}
         \sigma_{L^{'}}^2 & \mu_3^{'} \\
         \mu_3^{'} & \mu_4^{'}-\sigma_{L^{'}}^4 \\
       \end{array}
     \right)
     \bm{\dot{\textbf{g}}}(\mu_{L^{'}},\sigma_{L^{'}}^2) \notag \\
&=\frac{\pi^2 e^{ 2( \mu_{L^{'}} + \mathbb{C})}\left( \alpha^2 + 4\right)}{12 \alpha^2}\notag \\
&=\frac{\pi^2 \gamma^2\left(\alpha^2+ 4 \right)}{12 \alpha^2}, \notag
\end{align}
and the last line is obtained by substituting   $( \log ( \gamma ) - \mathbb{C} )$ for  $\mu_{L^{'}}$.  Similarly,
\begin{align}
\sqrt{n}\left(\hat{\alpha}-\alpha\right)&
\stackrel{d}{\longrightarrow} \textsl{N} \left[0, \;
\left(\frac{ \pi }{2 \sqrt{3} \left(\sigma_{L^{'}}^2-\pi^2/12
\right)^{3/2}} \right)^2  \left(\mu_4^{'}-\sigma_{L^{'}}^4 \right) \right] \notag \\
& = \textsl{N} \left[0, \;
\frac{\alpha^2}{80}\left( 13 \alpha^4 + 20 \alpha^2 + 64\right)\right],\notag
\end{align}
where the final simplification is attained by plugging in $\sigma_{L{'}}^2=\frac{\pi^2 (\alpha^2+4)}{12 \alpha^2}$ (see \ref{3.2}),
\[
\textbf{g}(\mu_{L^{'}},\sigma_{L^{'}}^2) = \frac{\pi}{\sqrt{3\left( \sigma_{L^{'}}^2 -\pi^2/12 \right)}},
\]
and
\[
\bm{\dot{\textbf{g}}}(\mu_{L^{'}},\sigma_{L^{'}}^2)= \left(
                                         \begin{array}{c}
                                           0 \\
                                         \pi \bigg/  \left(2 \sqrt{3} \left(\sigma_{L^{'}}^2-\pi^2/12
\right)^{3/2}\right)    \\
                                         \end{array}
                                        \right).
\]
Therefore, we have shown that our method-of-moments estimators are
normally distributed (asymptotically unbiased) as the sample size $n$ goes large. Consequently, we can now
approximate the $(1-\varepsilon)100\%$ confidence interval for  $\alpha$ and $\gamma$ as
\[
\hat{\alpha} \pm z_{\varepsilon/2}\sqrt{\frac{\hat{\alpha}^2 \left( 13 \hat{\alpha}^4 + 20 \hat{\alpha}^2 + 64 \right)}{80 n}},
\]
and
\[
\hat{\gamma} \pm
z_{\varepsilon/2}\sqrt{\frac{\pi^2 \hat{\gamma}^2\left(\hat{\alpha}^2+ 4 \right)}{12 \hat{\alpha}^2 n}},
\]
respectively, where $z_{\varepsilon/2}$ is the $(1-\varepsilon/2)$th quantile of the standard normal distribution, and $0 <\varepsilon<1$.


\section{Testing   MoM estimators on  simulated data}

In this section, we computationally test the MoM estimators of $\alpha$ and $\gamma$ obtained in Section 4 using the median absolute deviation (MAD) from the true values of our parameters as our criterion.  We generated $2000$ samples of  sizes $n = 100, 1000$, and 10000. The  estimates $\hat \alpha$ and $\hat \gamma$   for each of the $m$ samples and the average are then calculated. These values are shown in Table 1 below. When the sample size is at least $n =10000$, the relative fluctuations of the estimates of $\alpha$  and $\gamma$ are  becoming less than $2.2\%$ and $17.7\%$, respectively. Notice that the estimator of  $\gamma$ has a large variation when $\alpha$ is close to 0. It is also seemingly biased for both small values of $\alpha$ and small sample sizes. Nevertheless, it can be seen from Table 1 that the point estimators are asymptotically unbiased as expected.


\begin{table}[h!t!b!p!]
\caption{\emph{Mean estimates of and  dispersions from the true parameters  $\alpha$ and $\gamma$.}} \centerline {
\begin{tabular*}{4.4in}{@{\extracolsep{\fill}}|c||c@{\hspace{0.1in}}|cc|cc|cc|}
 \hline
\multirow{2}{*}{$(\alpha, \gamma)$} &  &  \multicolumn{2}{c|}{$n=100$}  &  \multicolumn{2}{c|}{$n=1000$} &  \multicolumn{2}{c|}{$n=10000$}  \\
 & & Mean & MAD& Mean& MAD & Mean & MAD \\
  \hline \hline
\multirow{2}{*}{$(0.1, 0.05)$}  &  $\alpha$     & 0.101  & 0.009    &0.100  & 0.003 &  0.100 & 0.001  \\
               &  $\gamma$     & 0.250  & 0.071    &0.058  & 0.027  & 0.051 & 0.009  \\
\hline
\multirow{2}{*}{$(0.2, 0.5)$}  &  $\alpha$    & 0.202  & 0.017    &0.200  & 0.006 &  0.200 & 0.002  \\
              &  $\gamma$     & 0.736  & 0.400    &0.518  & 0.141  & 0.502 & 0.048  \\
\hline
\multirow{2}{*}{$(0.5, 1000)$}  &  $\alpha$    & 0.507  & 0.047          &0.501    & 0.014 &  0.500 & 0.004  \\
              &  $\gamma$     & 1056.029   & 344.901    &1007.287  & 116.633  & 1000.422 & 35.313  \\
  \hline
\multirow{2}{*}{$(0.8, 100)$}  &  $\alpha$    & 0.811  & 0.082       &0.801     & 0.026 &  0.800 & 0.008  \\
              &  $\gamma$     & 103.865  & 24.042    &100.382  & 7.803  & 100.021 & 2.455  \\
  \hline
\multirow{2}{*}{$(1, 0.2)$}  &  $\alpha$      & 1.018  & 0.107    &1.001  & 0.035 &  1.000 & 0.010  \\
              &  $\gamma$     & 0.203  & 0.039    &0.200  & 0.012  &0.200 & 0.004  \\
  \hline
\multirow{2}{*}{$(1.2, 10)$}  &  $\alpha$      & 1.234  & 0.146     &1.202  & 0.045 &  1.200 & 0.014  \\
              &  $\gamma$     & 10.162  & 1.757    &10.023  & 0.581  & 9.997 & 0.179  \\
  \hline
\multirow{2}{*}{$(1.75, 1)$} &  $\alpha$        & 1.845 & 0.317    &1.760  & 0.096 &  1.750 & 0.031  \\
              &  $\gamma$     & 1.011 & 0.142    &1.000  & 0.041  & 1.000 & 0.013  \\
  \hline
\multirow{2}{*}{$(2, 0.1)$} &     $\alpha$         & 2.170  & 0.429   &2.014  & 0.130     & 2.000 & 0.042  \\
              & $\gamma$     & 0.101 &  0.012    &0.100  & 0.004  & 0.100 & 0.001  \\
  \hline
\end{tabular*}
}
  \label{t1}
\end{table}

In the interval calculations, we simulated 2000 sets of sample size   $n$ and averaged the lower and upper $95\%$ confidence bounds using the formula obtained in Section 4. Table \ref{t2}  below shows the   asymptotic behavior of the confidence intervals using sample sizes $n=100, 1000,$ and 10000.  Generally, it can  be seen  that  the asymptotic interval estimators performed quite satisfactorily for different combinations of the parameter values. We emphasize that one can always use bootstrap methods as the explicit forms of the estimators are known. But the asymptotic-based interval estimates are faster and easier to compute.

\begin{table}[h!t!b!p!]
\caption{\emph{Average 95\% confidence intervals  for different values of $\alpha$ and  $\gamma$.}}
 \begin{small}
 \centerline {
\begin{tabular*}{4.4in}{|c@{\hspace{0.05in}}||c@{\hspace{0.05in}}|c@{\hspace{0.05in}}|c@{\hspace{0.05in}}|c@{\hspace{0.05in}}|}
 \hline
 &  &  $n=100$  & $n=1000$ &  $n=10000$  \\
  \hline \hline
\multirow{2}{*}{$(0.1, 0.05)$}  &  $\alpha$    & (0.083 , 0.118))  & (0.095 ,  0.105)  & (0.098 ,  0.101)   \\
                &  $\gamma$    & (-0.733 , 1.297)   &(-0.008 , 0.127)  & (0.033 ,  0.068)  \\
\hline
\multirow{2}{*}{$(0.2, 0.5)$}  &  $\alpha$    & (0.167 , 0.239)   &  (0.189 , 0.211)   & (0.196 , 0.203)  \\
              &  $\gamma$     &(-0.588 , 2.086)  &  (0.213 , 0.830)   & (0.412 , 0.591) \\
\hline
\multirow{2}{*}{$(0.5, 1000)$}  &  $\alpha$    & (0.411 , 0.596)       & (0.471 , 0.529)        & (0.491 , 0.509)  \\
              &  $\gamma$     & (290.060 , 1844.721)  & (772.112 , 1237.549)    & (926.761 , 1073.340)  \\
  \hline
\multirow{2}{*}{$(0.8, 100)$}  &  $\alpha$    & (0.649 , 0.974)     &  (0.750 , 0.851)       &  (0.785 , 0.816) \\
              &  $\gamma$    & (54.162 , 152.626)  & (85.195 , 115.581)    & (95.334 , 104.913) \\
  \hline
\multirow{2}{*}{$(1, 0.2)$}    &  $\alpha$    & (0.796 , 1.247)  & (0.932 , 1.068)     & (0.978 , 1.021)   \\
              &  $\gamma$    & (0.124 , 0.285)  & (0.175 , 0.225)     &  (0.192 , 0.208)  \\
  \hline
\multirow{2}{*}{$(1.2, 10)$}  &  $\alpha$    & (0.920 , 1.529)  & (1.111 , 1.293)     & (1.171 , 1.228)   \\
              &  $\gamma$    &(6.644 , 13.594)  & (8.916 , 11.101)    & (9.654 , 10.344) \\
  \hline
\multirow{2}{*}{$(1.75, 1)$}   &  $\alpha$        & (1.083 , 4.003)  & (1.563 , 1.949)     & (1.691 , 1.811)     \\
              &  $\gamma$        & (0.733 , 1.268)   &  (0.915 , 1.085)    & (0.973 , 1.026)  \\
  \hline
\multirow{2}{*}{$(2, 0.1)$}    &  $\alpha$     & (0.925 , 3.448)  &  (1.748 , 2.282)   &  (1.920 , 2.085)       \\
              &  $\gamma$     & (0.076 , 0.126) & (0.092 , 0.108)    & (0.098 , 0.102)   \\
                \hline
\end{tabular*}
}
\end{small}
  \label{t2}
\end{table}


Table 3 shows the corresponding coverage probabilities of the interval estimates above. Apparently, the proposed interval estimator of $\alpha$  performed relatively well  near the boundaries for sample size $n=100$. Note that \cite{jrt99}'s interval estimator performs satisfactorily only when $\alpha <1$ even with sample size $n=500$. In addition, the table below insinuates that if the true parameter value of  $\alpha$  is close to 0 then we need at least 1000 observations to obtain reasonable estimates for small values of $\gamma$. Nonetheless, the coverage probabilities still provide relatively good merits for our estimators especially for sample size $n=100$.

\begin{table}[h!t!b!p!]
\caption{\emph{Coverage probabilities of 95\% interval estimates  for different values of $\alpha$ and  $\gamma$.}}
 \begin{small}
 \centerline {
\begin{tabular*}{2.7in}{|c@{\hspace{0.05in}}||c@{\hspace{0.05in}}|c@{\hspace{0.05in}}|c@{\hspace{0.05in}}|c@{\hspace{0.05in}}|}
 \hline
 &  &  $n=100$  & $n=1000$ &  $n=10000$  \\
 \hline \hline
 \multirow{2}{*}{$(0.1, 0.05)$}  &  $\alpha$    & 0.954   & 0.948  & 0.949   \\
                &  $\gamma$    & 0.803   & 0.914  &  0.947  \\
\hline
\multirow{2}{*}{$(0.2, 0.5)$}  &  $\alpha$     &0.945   & 0.943   & 0.954  \\
              &  $\gamma$     &0.859   & 0.944   & 0.955  \\
\hline
\multirow{2}{*}{$(0.5, 1000)$}  &  $\alpha$    & 0.959      & 0.952    & 0.948  \\
              &  $\gamma$     & 0.933      & 0.945    & 0.946  \\
  \hline
\multirow{2}{*}{$(0.8, 100)$}  &  $\alpha$    & 0.961   & 0.956      & 0.951 \\
              &  $\gamma$    & 0.944   & 0.945      & 0.946 \\
  \hline
\multirow{2}{*}{$(1, 0.2)$}    &  $\alpha$    & 0.953  & 0.949     & 0.955   \\
              &  $\gamma$    & 0.949  & 0.938     & 0.948  \\
  \hline
\multirow{2}{*}{$(1.2, 10)$}  &  $\alpha$    & 0.958  & 0.942    & 0.943   \\
              &  $\gamma$    &0.943  & 0.949    & 0.946  \\
  \hline
\multirow{2}{*}{$(1.75, 1)$}   &  $\alpha$     & 0.942  & 0.956    & 0.955     \\
              &  $\gamma$     & 0.951  & 0.946    & 0.953  \\
  \hline
\multirow{2}{*}{$(2, 0.1)$}    &  $\alpha$     & 0.940 &  0.959   & 0.947    \\
              &  $\gamma$     & 0.946 &  0.949   & 0.943    \\
  \hline
 \multirow{2}{*}{Average}       &  $\alpha$     & \bf{0.951}  & \bf{0.950}   & \bf{0.950}       \\
              &  $\gamma$      & \bf{0.916}  & \bf{0.941}  & \bf{0.948} \\
  \hline
\end{tabular*}
}
\end{small}
  \label{t3}
\end{table}

We now attempt to compare our procedure with that of \cite{jrt99}  and \cite{koz01} by using the mean point estimate (unbiasedness for finite samples) and  coefficient of variation (CV) as criteria.  Additionally, we  followed \cite{koz01} by focusing on $\alpha \in (1,2]$ for simplicity and due to the likely applications in finance in this range. We utilized the same constants $q_1=0.5, q_2=1$ of \cite{koz01},  and  $\lambda_1=0.001, \lambda_2=0.1$  of  \cite{jrt99}. The results are showcased in Table 4 below using  2000 samples of size $n=100, 1000, 10000$ each. It can be seen that the maximum CV of the proposed estimators is 26.85\% while \cite{jrt99}  and \cite{koz01} have 44.05\% and 40.11\%, correspondingly. Undoubtedly, the proposed procedure outperformed  the competing methods in estimating $\gamma$. It also performed  best in estimating $\alpha$ except when the true value is close to 2 and the sample size $n<10000$. Likewise, we generated 10000 samples of size $n=100$, and obtained the coverage probabilities 95.1\% and 91.7\% for $\alpha=0.4$ and $\gamma=0.5$, respectively compared with the 93.1\% and 86.5\% of \cite{jrt99}. These consequences considerably provide an additional argument in favor of the proposed approach.  On the contrary, it  appears that \cite{jrt99}'s procedure needs further tuning of the $\lambda$ values to obtain better results.

\begin{table}[h!t!b!p!]
\caption{\emph{Comparison of point estimators  for different values of $\alpha$,  $\gamma$, and sample size $n$.}}
 \begin{small}
 \centerline {
\begin{tabular*}{4.63in}{|c@{\hspace{0.05in}}|c@{\hspace{0.09in}}|c@{\hspace{0.09in}}||c@{\hspace{0.09in}}|c@{\hspace{0.09in}}|c@{\hspace{0.09in}}||c@{\hspace{0.09in}}|c@{\hspace{0.09in}}|c@{\hspace{0.09in}}|}
 \hline
$(\alpha, \gamma)$ & $n$ &  &  $\hat{\alpha}$  & $\hat{\alpha}_K$ &  $\hat{\alpha}_P$ &  $\hat{\gamma}$  & $\hat{\gamma}_K$ &  $\hat{\gamma}_P$  \\
 \hline \hline
\multirow{6}{*}{$(1.1, 0.9)$} & \multirow{2}{*}{100}   &  Mean      & 1.125   & 1.328  & 1.632   & 0.919   & 1.142  & 1.927 \\
               &      &  CV(\%)    & 11.82   & 10.18 &  22.91   & 18.81   & 40.11  & 40.18 \\ \cline{2-9}
               &   \multirow{2}{*}{1000}  &  Mean     & 1.104   & 1.228  & 1.290   & 0.900   & 1.037  & 1.285 \\
               &       &  CV(\%)   & 3.64    & 6.34  &  22.33   & 5.89    & 8.74   & 44.05 \\ \cline{2-9}
               & \multirow{2}{*}{10000}  &  Mean    & 1.100   & 1.180  & 1.126   & 0.900   & 0.993  & 0.960 \\
               &      &  CV(\%)    & 1.16   &  4.37   &  9.57   & 1.87    & 9.57   & 22.63 \\
 \hline
\multirow{6}{*}{$(1.3, 2)$ }    &  \multirow{2}{*}{100} &  Mean      & 1.341   & 1.468  & 1.699    & 2.038   & 2.223  & 2.861 \\
               &      &  CV(\%)    & 13.99   & 10.78  & 18.92   & 16.94   & 17.13  & 29.61 \\ \cline{2-9}
               & \multirow{2}{*}{1000}  &  Mean     & 1.301   & 1.374  & 1.468   & 2.002   & 2.103  & 2.382 \\
               &       &  CV(\%)   & 4.14    & 6.62   & 16.57   & 5.16    & 6.14    & 24.99 \\ \cline{2-9}
               & \multirow{2}{*}{10000}  &  Mean    & 1.300   & 1.338  & 1.334   & 2.000   & 2.054  & 2.077 \\
               &      &  CV(\%)    & 1.29    & 4.39  &  8.33    & 1.62    & 4.22   & 13.15 \\
 \hline
\multirow{6}{*}{$(1.5, 1)$} &  \multirow{2}{*}{100}     &  Mean      & 1.555   & 1.617  & 1.896   & 1.013   & 1.049  & 1.574 \\
               &      &  CV(\%)    & 16.00   & 10.12  & 11.17   & 15.45   & 14.38  & 27.23 \\ \cline{2-9}
               & \multirow{2}{*}{1000}  &  Mean     & 1.504   & 1.547  & 1.759   & 1.001  &  1.018  &  1.401 \\
               &       &  CV(\%)   & 4.72    & 5.97  &  12.64   & 4.81   &  5.31   &  24.01 \\ \cline{2-9}
               & \multirow{2}{*}{10000}  &  Mean(\%) & 1.500   &1.514  & 1.603   & 1.000   & 1.006  & 1.164 \\
               &      &  CV(\%)    &  1.51   & 3.74  &  11.6    & 1.51   &  3.07  &  24.14 \\
 \hline
\multirow{6}{*}{$(1.7, 10)$} &  \multirow{2}{*}{100}  &  Mean        & 1.791   & 1.779  & 1.836   & 10.127   & 10.240  & 10.077 \\
               &      &  CV(\%)    & 20.36   & 8.43  &  8.85     & 14.61   & 13.04  & 13.91 \\ \cline{2-9}
               & \multirow{2}{*}{1000}  &  Mean     & 1.707   & 1.721  & 1.752   & 10.005   & 10.045  & 9.997 \\
               &       &  CV(\%)   & 5.52    & 5.09  &  7.30    & 4.52     & 4.59  & 4.51 \\ \cline{2-9}
               & \multirow{2}{*}{10000}  &  Mean    & 1.701   & 1.705  & 1.710   & 9.995    & 10.005  & 9.995 \\
               &      &  CV(\%)    & 1.70    & 2.79 &   3.70   &  1.38     & 2.04  & 1.41 \\
\hline
\multirow{6}{*}{$(1.99, 0.1)$} &  \multirow{2}{*}{100}  &  Mean       & 2.129   & 2.010  & 2.000   & 0.101   & 0.101  & 0.100 \\
               &      &  CV(\%)    & 26.85   & 5.52  &  0.00    & 12.56   & 11.11  & 11.89 \\ \cline{2-9}
               & \multirow{2}{*}{1000}  &  Mean     & 1.998   & 1.990  & 1.999   & 0.100   & 0.100  & 0.102 \\
               &       &  CV(\%)   & 6.78    & 2.01  &  0.13    &  4.15   & 3.64  &  12.05 \\ \cline{2-9}
               & \multirow{2}{*}{10000}  &  Mean    & 1.991   & 1.990  & 1.999   & 0.100   & 0.100  & 0.101 \\
               &      &  CV(\%)    & 2.06   &  0.67  &  0.93    & 1.29    & 1.15   & 3.82 \\
\hline
\end{tabular*}
}
\end{small}
  \label{t4}
\end{table}

Overall, Tables \ref{t1}-\ref{t4} strongly indicate that  the proposed point and interval  estimators did fairly well in our simulations. The result of the comparison further added merits to our method. The point estimates could be regarded as reasonable starting values for better iterative estimation algorithms too.

\section{Concluding remarks}

We have derived  the first few moments of a log-transformed Linnik distributed random variable. The derivation led to a system of estimating equations which are then used to estimate the parameters of the L$(\alpha, \gamma)$ distribution.  A major advantage of our estimators is that we do not need to choose appropriate constants a priori to obtain reasonable parameter estimates. We emphasized that the pre-selection of these constants is not straightforward in practice. Furthermore, the proposed estimators are computationally simpler and faster as we do not need to  solve a system of non-linear equations using iterative numerical methods.  In general, the proposed estimators outperformed the competing procedures.

To conclude, we cite a few extensions which would be worth pursuing in the future. For instance, improving the above estimators by bootstrapping, and  developing other estimators using the likelihood  approach would be of interest. The derivation of the corresponding method-of-moments estimators for the multivariate case as in \cite{a92,oa08,kak06}, and  the  application of our procedure in practice particularly in finance and economics would be valuable pursuits as well.

%% file: CahoyMain.bbl
\begin{thebibliography}{}

\bibitem{a92}Anderson, D.N., A multivariate Linnik distribution, Statistics \& Probability Letters, 14(4), 333--336 (1992).

\bibitem{aaa93}Anderson, D.N.,   Arnold, B.C., Linnik Distributions and Processes, Journal of Applied Probability, 30(2), 330--340 (1993).

\bibitem{oa08}Arslan, O., An alternative multivariate skew Laplace distribution: properties and estimation, Statistical Papers, 30(2), DOI: 10.1007/s00362-008-0183-7 (2008).

\bibitem{bkkuz04} Bening, V.E., Korolev, V.Y., Kolokol'tsov, V.N., Saenko, V.V., Uchaikin, V.V., Zolotarev, V.V.,  Estimation of parameters of fractional stable distributions. Journal of Mathematical Sciences,  123(1), 3722--3732 (2004).

\bibitem{cuw10}Cahoy, D.O., Uchaikin, V.V., Woyczynski, W.A.,  Parameter estimation in fractional Poisson processes. Journal of Statistical Planning and Inference,  140(11), 3106--3120 (2010).

\bibitem{dev90}Devroye, L.,  A note on Linnik's distribution. Statistics \& Probability Letters, 9, 305--306 (1990).

\bibitem{fer96} Ferguson, T., A Course in Large Sample Theory, Chapman \& Hall, (1996).

\bibitem{jrt99}Jacques, C., R$\acute{e}$millard, B.,  Theodorescu, R. Estimation of Linnik law parameters, Statistics \& Decisions, 17(3), 213--236 (1999).

\bibitem{hnr80}Haan, L.D., Resnick, S.I.,  A simple asymptotic estimate for the index of a stable distribution,  Journal of the Royal Stat'l Soc B, 42, 83-87 (1980).

\bibitem{hall82}Hall, P. On some simple estimates of an exponent of regular variation, J. Roy. Statist. Soc. Ser. B., 44, 37--42 (1982).

\bibitem{hil75}Hill, B.M., A simple general approach to inference about the tail of a distribution,  Annals of Statistics, 3, 1163-1174 (1975).

\bibitem{koz99}Kozubowski, T.J.,  Geometric stable laws: estimation and applications, Mathematical and Computer Modelling, 241--253 (1999).

\bibitem{koz01}Kozubowski, T.J.,  Fractional moment estimation of Linnik and Mittag-Leffler parameters, Mathematical and Computer Modelling, 1023--1035 (2001).

\bibitem{kao96}Kotz, S.,  Ostrovskii, I.V., A mixture representation of the Linnik distribution, Statistics \& Probability Letters, 26(1),  61--64 (1996).

\bibitem{kak06}Kuttykrishnan, A.P., K. Jayakumar, K., Bivariate semi $\alpha$-Laplace distribution and processes,  Statistical Papers, 49(2), 303--313 (2006).

\bibitem{lap75}Leitch, R.A., Paulson, A.S., Estimation of stable law parameters: stock price behavior application. J. Amer. Statist. Assoc. 70, 690--697 (1975).

\bibitem{lin98a}Lin, G.D., A note on the Linnik distributions, Journal of Mathematical Analysis and Applications, 217, 701--706 (1998).

\bibitem{lin63}Linnik, J.V., Linear forms and statistical criteria. I.  II, Select Translat Math. Statist. Probab., 3, 1--90 (1963).

\bibitem{pake98}Pakes, A.G., Mixture representations for symmetric generalized Linnik laws, Statistical Probability Letters, 37, 213--221 (1998).

\bibitem{phl75} Paulson, A.S., Holcomb, E.W., Leitch, R.A.,  The estimation of the parameters of the stable laws. Biometrika 62, 163--170 (1975).

\bibitem{press72}Press, R.N. Estimation in univariate and multivariate stable distribution, J. Amer. Statist. Assoc., 67, 842--846 (1972).

\bibitem{zol86} Zolotarev, V.M., One-dimensional Stable Distributions: Translations of Mathematical Monographs,  Vol 65, American Mathematical Society, United States of America, (1986).

\end{thebibliography}
